\documentstyle[12pt,aaspp]{article}
\begin{document}
\title{An Incoherent $\alpha-\Omega$ Dynamo in Accretion Disks}
\author{Ethan T. Vishniac}
\affil{Department of Astronomy, University of Texas, Austin TX 78712, USA\\
I: ethan@astro.as.utexas.edu}
\author{Axel Brandenburg}
\affil{Nordita, Blegdamsvej 17, DK-2100, Copenhagen \O,
Denmark I:brandenb@nordita.dk}
\begin{abstract}
We use the mean-field dynamo equations to show that an incoherent
alpha effect in mirror-symmetric turbulence in a shearing
flow can generate a large scale, coherent magnetic field.
We illustrate this effect with simulations of a few simple
systems.  In accretion disks,
this process can lead to axisymmetric magnetic domains whose radial
and vertical dimensions will be comparable to the disk height.
This process may be responsible for observations of
dynamo activity seen in simulations of
dynamo-generated turbulence involving, for example,
the Balbus-Hawley instability.
In this case the magnetic field strength will saturate at $\sim (h/r)^2$
times the ambient pressure in real accretion disks.
The resultant dimensionless viscosity will be
of the same order.  In numerical simulations the azimuthal extent
of the simulated annulus should be substituted for $r$.  We compare
the predictions of this model to numerical simulations previously
reported by Brandenburg et al. (1995).  In a radiation
pressure dominated environment this estimate for viscosity should be reduced
by a factor of $(P_{gas}/P_{radiation})^6$ due to magnetic buoyancy.
\end{abstract}
\section{Introduction}

Understanding the transport of angular momentum in accretion disks
is one of the basic challenges in modern astrophysics.  The
traditional approach (\cite{ss73}) is to assume that accretion
disks are characterized by an effective viscosity, arising
from an unspecified collective process, given by $\alpha_{SS} c_s h$, where
$c_s$ is the local sound speed, $h$ is the disk half-thickness,
and $\alpha_{SS}$ is a constant of order unity.  More recently, there
has been the realization (\cite{bh91}) that a previously discovered
magnetic field
instability in a shearing flow (\cite{v59}, \cite{c61}) will
act to produce a positive angular momentum flux in an accretion
disk.  This has given rise to two separate, but
related claims.  The first is the proposal that this is the
dominant mechanism of angular momentum transport in ionized
accretion disks.  The second is the proposal that this instability,
by itself, leads to a turbulent dynamo which drives the magnetic
field into equipartition with the ambient pressure, i.e.
$V_A\sim c_s$, where $V_A$ is the Alfv\'en speed in the disk.
The growth rate for this dynamo is usually taken, following the
original claim of Balbus and Hawley, to be $\sim\Omega$.
Since the dimensionless `viscosity', $\alpha_{SS}$, is $\sim (V_A/c_s)^2$,
this proposal requires that $\alpha_{SS}$ be a number `of order
unity'.  However, we need to interpret this phrase generously.
In numerical simulations (e.g. \cite{bnst95}) $\alpha_{SS}$
is less than $10^{-2}$ both because the magnetic pressure saturates
at a fraction of the gas pressure, and because the off-diagonal
components of $\langle\vec B\vec B\rangle$ are a fraction of $B_\theta^2$.

Three dimensional simulations of the Balbus-Hawley instability
have been performed by a variety of researchers, with and without
imposed vertical magnetic flux, and with and without vertical
disk structure (\cite{hgb95a}, \cite{bnst95}, \cite{shgb95},
and \cite{hgb95b}).  We note in particular Brandenburg et al.
(1995) in which no net flux was imposed on the computational box,
and vertical disk structure was included.  In this simulation,
as in the others, there was an
initial rise in the magnetic energy density at a rate $\sim\Omega$.
At the end of this phase the system had not yet lost memory of
its initial conditions, but after a somewhat longer time,
which may be as long as a few dozen rotation periods, the
simulation asymptotically approached a final state with $V_A\sim c_s$.
The approach to this state was characterized by the appearance of
a large scale field which underwent
spontaneous reversals at irregular intervals of tens of rotational
periods.

Interestingly, the presence of a large scale coherent field
does not seem to be due to an $\alpha-\Omega$ dynamo, because
the relative helicity is just a few percent.
Conventional $\alpha-\Omega$
dynamo models rely on a nonzero $\alpha_{\theta\theta}$
component in the helicity tensor (not to be confused with the
dimensionless viscosity, written here as $\alpha_{SS}$) to
produce a large scale coherent field.  The presence of an initial rapid rise
is less surprising, since imposing a uniform large scale magnetic field
in a turbulent medium results in the formation of intermittent
magnetic field structures and a consequent rise in the magnetic
energy density at the turbulent eddy turn-over rate.  In addition,
there is evidence (\cite{mfp81}) that turbulence in a conducting
fluid can generate a modest and highly disordered magnetic field
even in the absence of an imposed global field.  Both of these
effects are probably due to the ability of of symmetric turbulence to
produce a negative effective diffusion coefficient (\cite{m78}) and they
both can be relied open to contribute to the growth of the high wavenumber
component of the magnetic field.  On the other hand,
the slower relaxation rate seen after the initial rise is correlated
with changes in the large scale field and is presumably an indicator
of the large scale dynamo growth rate.  Since the turbulence is
sustained by an instability of the large scale field, its ability
to generate such a field is critically important.

The saturation level of the magnetic field in these simulations
also leads to some puzzling questions.
The claim that the Balbus-Hawley instability
saturates when the magnetic pressure is comparable to the
ambient thermal pressure, and that the dimensionless
viscosity has some approximately fixed value, is difficult to
reconcile with attempts to model accretion disks in compact
binary systems.  Successful models of dwarf novae outbursts and
X-ray transients (\cite{s84a}, \cite{s84b}, \cite{mmh84}, \cite{hw89},
\cite{mw89a}, and, more recently, \cite{c94}),
as well as the distribution of light in quiescent
dwarf novae disks (\cite {mw89}) all imply that the dimensionless
viscosity, $\alpha_{SS}$, varies spatially and with time.  These variations
are consistent with $\alpha_{SS}\propto (h/r)^n$, where $n$ is a
constant lying somewhere between $1$ and $2$.
Recent work (\cite{ccl95}) on X-ray transients suggests that $n$ may
be close to $1.5$.  Here we note only that any value of $n$
appreciably different from zero conflicts
with claims for a universal value of $\alpha_{SS}$.

This difficulty can be resolved in several different ways.
{}For example, we might claim that magnetic instabilities
dominate $\alpha_{SS}$ only at low temperatures and that some
other process, e.g. convection, dominates at higher
temperatures.  This idea faces two major objections.
{}First, it explains only some of the phenomenological evidence
favoring a varying $\alpha_{SS}$.  Second, attempts to model
the vertical structure of dwarf novae disks invariably
conclude that such disks are convective during quiescence,
when $\alpha_{SS}$ is small and stably stratified during
outburst, when $\alpha_{SS}$ is relatively large (for a recent
discussion of the conditions necessary for convection
in partially ionized accretion disks see Cannizzo
1992).  This implies that convection could explain the
rise in $\alpha_{SS}$ only if it acts to suppress angular
momentum transport, rather than enhance it.  Alternatively,
one could appeal to the temperature dependence of the
resistivity to account for this effect, although the
effective resistivity of the simulations is, in any
case, many orders of magnitude larger than in real
disks.  A more promising notion is that one
might ascribe the rise in $\alpha_{SS}$ to the greater
thermal conductivity of disk in the hot state, although
the rationale for this is not yet clear.
{}Finally, one might simply conclude that
all the phenomenological models are wrong, for a variety
of reasons, a viewpoint which is difficult to dismiss
given the large uncertainties faced in modeling
accretion disks.

In this paper we will explore a new turbulent disk
dynamo in which the turbulence is
not assumed to lack mirror symmetry in the vertical direction.
The dynamo effect arises from the fact that in a system
of finite size the mean square helicity, and the
instantaneous spatially averaged helicity, is still nonzero.
We will see that this leads to the existence of a
modified $\alpha-\Omega$ dynamo, in which the large
scale organization of the magnetic field comes from
the existence of a large scale shear.  Of course,
a real disk has vertical structure, which breaks the
vertical symmetry and allows for the {\it possibility} of
a nonzero average helicity.  However, we will show that
the incoherent dynamo mechanism will be particularly effective
in simulations of limited azimuthal extent.
Indeed the simulations of Brandenburg et al. (1995) show that
while the relative helicity is small (less than a few percent),
there is still a dynamo effect leading to the generation of
large scale fields.
Moreover, the saturation level of the magnetic field,
and the consequent value of $\alpha_{SS}$, turn out to
depend sensitively on the ratio $h/r$.  If this
dynamo is the only effect to arise from the turbulence
induced by the Balbus-Hawley effect, then it is
relatively simple to reconcile the dynamo activity
seen in simulations with phenomenological models
of accretion disks in compact binary systems.

In \S 2 we discuss the conceptual basis of an incoherent dynamo
in a turbulent shearing medium and estimate the growth rate.  In
\S 3 we apply this to accretion disks and show that the incoherent
dynamo gives a positive growth rate only for axisymmetric magnetic
domains.  We estimate the saturated state of the field and discuss
our results in light of numerical simulations of magnetic fields in a
Keplerian disk.
In \S 4 we summarize our results and their implications for astrophysical
disks and numerical simulations of such disks.

\section{The Incoherent Dynamo}

In a highly conducting medium the magnetic field obeys the induction
equation
\begin{equation}
\partial_t \vec B=\vec\nabla\times(\vec V\times\vec B),
\label{eq:ind}
\end{equation}
where we have neglected ohmic diffusion.  Ultimately this term is
important in allowing reconnection and smoothing.  Here we assume
that these processes take place at a rate determined by turbulent
processes.
The usual approach to dynamo theory is to define the response of
the large scale magnetic field to small scale motions as $\vec b$
and to derive its effects on the large scale field by substituting
$\vec b$ back into the right hand side of equation (\ref{eq:ind}).
{}For an incompressible fluid this yields
\begin{equation}
\partial_t B_i=\epsilon_{ijk}\partial_j (\alpha_{kl} B_l)+
\partial_j D_{jk}\partial_k B_i- \partial_j D_{ik}\partial_k B_j,
\label{eq:dyn1}
\end{equation}
where $\epsilon_{ijk}$ is the Levi-Civita tensor and
\begin{equation}
\alpha_{kl}\equiv \langle \epsilon_{kij} V_i\partial_l\int^t V_j(t')dt'
\rangle,
\label{eq:alpha}
\end{equation}
and
\begin{equation}
D_{jk}=\langle V_j\int^t V_k(t')dt'\rangle.
\label{eq:diff}
\end{equation}
The first term comes from the stretching of large scale field lines by
the local turbulence.  The tensor $\alpha_{kl}$ describes the twisting
of large scale field lines into a spiral shape.  Reconnection between
adjacent spirals produces a large scale field component at right angles
to the original field line provided that either the degree of twisting
or the large scale magnetic field strength varies in the third direction.
The second term is the usual diffusion term, modified by the presence
of the third term.

We can see from equation (\ref{eq:alpha}) that each component of
$\alpha_{ij}$ has either a factor of $V_z$ or $\partial_z$.  If the
local velocity field is mirror symmetric, in the sense that its
statistical properties are unchanged under the transformation
$z\rightarrow -z$, then the time and space averaged value of
$\alpha_{ij}$ vanishes.  This poses a significant but, as we will see,
not insurmountable, obstacle to a successful dynamo.

Another problem is that equations (\ref{eq:dyn1}), (\ref{eq:alpha})
and (\ref{eq:diff}) are usually defined
kinematically, i.e. the velocity field is assumed to be imposed
on the magnetic field.  Once the magnetic field becomes sufficiently
powerful it will modify the flow, which is usually taken into
account by including a correction term proportional to $B^2$.
However, in a Keplerian shearing flow the magnetic field will
be unstable and the resulting turbulence will be directly correlated
with the magnetic field.  Nevertheless, as long as we define
$\vec V$ in terms of the motion of the magnetic field lines
equation (\ref{eq:dyn1}) will remain valid, if difficult to
solve.  Here we will define our results in terms of the properties
of $\alpha_{ij}$ and $D_{ij}$ regardless of their ultimate source.

In a Keplerian disk the dynamo equations can be simplified as
\begin{equation}
\partial_t B_r=-\partial_z(\alpha_{\theta\theta}B_\theta)-\partial_z
(V_b B_r)+\partial_z (D_{zz}\partial_z B_r),
\label{eq:br}
\end{equation}
and
\begin{equation}
\partial_t B_\theta=-{3\over 2}\Omega B_r-\partial_z(V_b B_\theta)+
\partial_z (D_{zz}\partial_z B_\theta)+\partial_r (D_{rr}\partial_r B_\theta),
\label{eq:bt}
\end{equation}
where $\Omega\propto r^{-3/2}$ is the rotation frequency, $V_b$ is
the buoyant velocity of the magnetic field lines relative to the
surrounding fluid, and $B_r$ and $B_\theta$ are the radial and azimuthal
components of the magnetic field.  Equations (\ref{eq:br}) and (\ref{eq:bt})
differ from equation (\ref{eq:dyn1}) in that we have allowed for the
presence of global shearing, and magnetic field line buoyancy.  In
addition, we have assumed that the diffusion matrix is diagonal, and
dropped the effects of helicity on the evolution of $B_\theta$, given
that the shearing of $B_r$ should dominate such effects.  Also, we
have retained only the $\alpha_{\theta\theta}$ term in equation
(\ref{eq:br}) since the critical feedback term in the dynamo equations
involves generating radial magnetic flux from the azimuthal component
of the field.  Finally, given that we are interested in applying these
equations to accretion disks whose thickness is a small fraction of
their radius, we have assumed that vertical gradients will dominate over
radial gradients.

Now let's assume that the turbulence is symmetric under $z\rightarrow -z$
so that $\langle \alpha_{\theta\theta}\rangle=0$.  Although this eliminates any
coherent helicity, the value of $\langle B_r^2\rangle$ can still increase
in a random walk.  Ignoring diffusion and buoyancy we see that the formal
solution for $B_r$ is
\begin{equation}
B_r=\int^t -\partial_z (\alpha_{\theta\theta}(t') B_\theta(t')) dt'.
\label{eq:tvar}
\end{equation}
Now by hypothesis, $\alpha_{\theta\theta}$ is uncorrelated over time
scales greater than some eddy correlation time $\tau_{eddy}$.  If the radial
magnetic
field is undergoing a random walk, then it will usually be far enough away
from zero that it will not change sign every eddy correlation time.  Since
$B_r$ drives $B_\theta$ through coherent shearing, this implies that the
correlation time for $B_r$ and $B_\theta$ is much greater than $\tau_{eddy}$.
Consequently, we can consider the integrand in equation (\ref{eq:tvar})
as consisting of a rapidly varying factor, $\alpha_{\theta\theta}$,
multiplying a slowly varying function.  Multiplying equation
(\ref{eq:tvar}) times equation (\ref{eq:br}) and ignoring diffusion
and buoyancy, as before, we see that the integral in equation
(\ref{eq:tvar}) is correlated with $\alpha_{\theta\theta}$ only over
the last eddy correlation time $\tau_{eddy}$.
Consequently, we can replace the integral
in the product with
$-\partial_z(\alpha_{\theta\theta}(t) B_\theta(t))\tau_{eddy}$.  This
implies
\begin{equation}
\partial_t\langle B_r^2\rangle\approx K_z^2
{\langle \hat\alpha_{\theta\theta}^2\rangle\over N}
\tau_{eddy}\langle B_\theta^2\rangle,
\label{eq:ran1}
\end{equation}
where $N$ is the number of independent turbulent eddies in a magnetic domain,
$K_z$ is the vertical wavenumber
of the magnetic domain, and $\langle \hat\alpha_{\theta\theta}^2\rangle$ is
the mean square helicity associated with a single eddy.  In general this
will be of order $V_T^2$ where $V_T$ is the root mean square turbulent
velocity.  Since $B_r$ is being driven incoherently we can expect it to
undergo frequent reversals.  In between such reversals the shearing of
the field will drive $B_\theta^2$  sharply upward. From equations (\ref{eq:bt})
and (\ref{eq:ran1}) we see that the correlation time of the radial
magnetic field and the growth time of the magnetic field are comparable and
given by
\begin{equation}
\tau_{corr}^{-1}\sim\tau_{growth}^{-1}\sim
  \left({K_z^2 V_T^2\Omega^2\tau\over N}\right)^{1/3}.
\label{eq:gr1}
\end{equation}
We note that $\tau_{corr}$ has to be greater than $\tau_{eddy}$ in
order for this estimate to be internally self-consistent, i.e. the
magnetic field must be correlated over longer times than the turbulence
itself.
Since a field reversal in $B_\theta$ requires that $B_r$ not
only reverse its sign, but maintain it long enough to push $B_\theta$
through zero, it is clear that the correlation time for $B_\theta$
may be somewhat larger than the correlation time for $B_r$.  We will
return to this point later.

By itself this argument does not show that a succession of random
twists in a shearing background can drive an exponential increase in
the magnetic field.  We need to show that the growth experienced
between field reversals dominates over the abrupt cancellation of
the field as $B_r$ reverses itself.  We also need to show that
our estimate of the growth rate given in equation (\ref{eq:gr1})
will dominate over turbulent diffusion for some range of magnetic
domain sizes.

We can test the assertion that a series of random changes in
$B_r$ can drive a dynamo by constructing a simple toy model of
the process, which ignores the spatial structure of the field,
but includes its  dynamical evolution.  Assuming that
$\alpha_{\theta\theta}$ has a stochastic component and ignoring
buoyancy
we can rewrite  equations (\ref{eq:br}) and (\ref{eq:bt}) as
\begin{equation}
\partial_t B_r=(\eta(t)-\alpha_{coh}) B_\theta - D B_r,
\label{eq:tbr}
\end{equation}
and
\begin{equation}
\partial_t B_\theta=-{3\over2}\Omega B_r-DB_\theta,
\label{eq:tbt}
\end{equation}
where $\eta(t)$ is a stochastic variable with a correlation
time $\tau_{eddy}$ and $\alpha_{coh}$ is the coherent component of
$\partial_z\alpha_{\theta\theta}$.
Here we have subsumed spatial derivatives into the definitions of
$\eta$ and $D$ and ignored the $-\alpha_{\theta\theta}\partial_zB_\theta$
term which would normally appear in the mean-field dynamo equations.
We have also assumed that turbulent damping is
the same for each component of the magnetic field, which is not generally
true, but simplifies the analysis without losing any essential
physics.  Equations (\ref{eq:tbr}) and (\ref{eq:tbt}) can be rewritten
in a more convenient form by defining $A\equiv (B_r/B_\theta)$.  Then
\begin{equation}
\partial_t A=\eta(t)-\alpha_{coh}+{3\over2}\Omega A^2,
\label{eq:stoch}
\end{equation}
and
\begin{equation}
\partial_t \ln B_\theta^2=-3\Omega A-2D.
\label{eq:growth}
\end{equation}
The magnetic field will grow exponentially if $\langle A\rangle$ is
negative and $-3\Omega\langle A\rangle>2D$.

We can find $\langle A\rangle$ by solving
equation (\ref{eq:stoch}) in terms of an unnormalized probability
distribution function $P(A)$ and evaluating
\begin{equation}
\langle A\rangle \equiv {\int_{-\infty}^{\infty} A P(A) dA\over
\int_{-\infty}^\infty P(A) dA}.
\end{equation}
The distribution function $P(A)$ satisfies the equation
\begin{equation}
\partial_A(\dot A P(A)-\langle\eta^2\rangle\tau_{eddy} \partial_A P(A))=0,
\end{equation}
or
\begin{equation}
P(A)\left({3\over 2}\Omega A^2-\alpha_{coh}\right)-
\langle\eta^2\rangle\tau_{eddy}\partial_A P(A)=\langle
\eta^2\rangle\tau_{eddy},
\label{eq:distev}
\end{equation}
where we have taken advantage of the unnormalized nature of $P(A)$ to
set the constant of integration to $\langle\eta^2\rangle\tau_{eddy}$.
Equation (\ref{eq:distev})
can be solved to yield
\begin{equation}
P(A)=\exp\left[{\Omega A^3-2\alpha_{coh}A\over
2\langle\eta^2\tau_{eddy}\rangle}\right]
\int_A^{\infty}\exp\left[{-\Omega r^3+2\alpha_{coh}r\over 2\langle
\eta^2\tau_{eddy}\rangle}\right]dr.
\label{eq:pdist}
\end{equation}
Consequently,
\begin{equation}
\langle
A\rangle=\left({2\langle\eta^2\tau_{eddy}\rangle\over\Omega}\right)^{1/3}
{\int_{-\infty}^\infty dy\int_y^\infty ds y\exp[y^3-s^3-\gamma(y-s)]
\over\int_{-\infty}^\infty dy\int_y^\infty ds\exp[y^3-s^3-\gamma(y-s)]},
\label{eq:expa}
\end{equation}
where
\begin{equation}
\gamma\equiv {\alpha_{coh}\over \langle\eta^2\tau_{eddy}\rangle}
\left({2\langle\eta^2\tau_{eddy}\rangle\over\Omega}\right)^{1/3}.
\end{equation}
Equation (\ref{eq:expa}) can be rewritten by defining new variables
$w\equiv y+s$ and $x\equiv s-y$ and integrating over $w$.  We
obtain
\begin{equation}
\langle A\rangle={-1\over2}\left({2\langle\eta^2\tau_{eddy}\rangle
\over\Omega}\right)^{1/3}
{\int_0^\infty x^{1/2}\exp[x(\gamma-{x^2\over4})]dx\over
\int_0^\infty x^{-1/2}\exp[x(\gamma-{x^2\over4})]dx}.
\label{eq:exact}
\end{equation}
When $\gamma$ is small we can expand $e^{w\gamma}\approx 1+\gamma w$
and obtain
\begin{equation}
\langle A\rangle \approx -0.32
({\langle\eta^2\tau_{eddy}\rangle\over\Omega})^{1/3}
(1+0.51\gamma),
\label{eq:res1}
\end{equation}
which implies that
\begin{equation}
\partial_t\ln B_\theta^2\approx 0.96 (\langle\eta^2\tau_{eddy}\rangle
\Omega^2)^{1/3}+0.61\alpha_{coh}
\left({\Omega\over\langle\eta^2\tau_{eddy}\rangle}\right)^{1/3}-2D.
\end{equation}
In other words, the magnetic field will grow exponentially roughly as
fast as the estimate given in equation (\ref{eq:gr1}).  This
will be suppressed by turbulent diffusion only when the damping rate
due to diffusion is comparable to the growth rate.

The existence of an incoherent dynamo emerges from the fact that the
distribution
function $P(A)$ given in equation (\ref{eq:pdist}) is biased towards
negative values of $A$.  This bias comes, paradoxically enough, from the
coherent, positive definite term in equation (\ref{eq:stoch}).  When
$A$ is sufficiently positive it evolves deterministically through $+\infty$
into negative values.  (Actually, $B_r$ doesn't change during this
phase.  This deterministic trajectory is merely a field reversal for
$B_\theta$.)  The end result is that whenever $A$ becomes large and
positive it rapidly switches to being large and negative.
Ultimately,
the sign of the bias is determined by the sign of $\partial_r\Omega$.
The frequency of such field reversals is given by examining the
probability distribution at large $A$ when the evolution of $P(A)$
is deterministic.  If we define $\tau(A)$ as the time it takes
for the field to move from some large positive value of $A$ to
$A=\infty$ then from equation (\ref{eq:stoch}) we see that
\begin{equation}
\tau(A)^{-1}={3\over 2}\Omega A,
\label{eq:tauA}
\end{equation}
where we have neglected $\alpha_{coh}$ since for $A$ sufficiently large
its effects can be ignored.  The field reversal rate is just the limit
of this rate times the statistical weight of the distribution between
$A$ and $\infty$.  In other words,
\begin{equation}
\tau_{rev}^{-1}=\lim_{A\rightarrow\infty}\tau(A)^{-1}{\int_A^\infty
P(s)ds\over\int_{-\infty}^{\infty} P(s) ds}.
\end{equation}
Substituting equations (\ref{eq:pdist}) and (\ref{eq:tauA}) into
this result, and making the change of variables, as before, to
$x$ and $w$ we have
\begin{equation}
\tau_{rev}^{-1}=\lim_{A\rightarrow\infty}{3\over2}\Omega A
{\int_{2A(\Omega/2\langle\eta^2\tau_{eddy}\rangle)^{1/3}}^\infty
\int_0^\infty
\exp\left[ w\left(\gamma-{1\over4}(w^2+3x^2)\right)\right]dw dx\over
\int_{-\infty}^\infty\int_0^\infty
\exp\left[ w\left(\gamma-{1\over4}(w^2+3x^2)\right)\right]dw dx}.
\end{equation}
Both the numerator and the denominator can be simplified by integrating
over $x$ to obtain
\begin{equation}
\tau_{rev}^{-1}=\left({3\over\pi}\right)^{1/2}{(2\langle\eta^2
\tau_{eddy}\rangle \Omega^2)^{1/3}\over \int_0^{\infty} w^{-1/2}
\exp\left[w(\gamma-w^2/4)\right]dw}.
\label{eq:trev}
\end{equation}
When $\gamma$ is small this becomes
\begin{equation}
\tau_{rev}^{-1}=0.53(\langle\eta^2\tau_{eddy}\rangle\Omega^2)^{1/3}
(1-0.53\gamma),
\end{equation}
i.e. a rate which is roughly half the e-folding rate for the
magnetic field energy.

When $\alpha_{coh}$ is large and positive we can evaluate the integrals
in equation (\ref{eq:exact}) by expanding around the maximum of
$x(\gamma-x^2/r)$.  We obtain
\begin{equation}
\langle A\rangle \approx -\left({2\alpha_{coh}\over 3\Omega}\right)^{1/2},
\end{equation}
so that
\begin{equation}
\partial_t\ln B_\theta^2\approx 2\left({3\over2}\alpha_{coh}\Omega\right)^{1/2}
-2D,
\end{equation}
which is the expected result for a coherent $\alpha-\Omega$ dynamo.
In this limit the magnetic field reversal rate becomes
\begin{equation}
\tau_{rev}^{-1}=0.68\alpha{coh}^{1/4}\langle\eta^2\tau_{eddy}\rangle^{1/6}
\Omega^{7/12} \exp
\left[{-1.1\alpha_{coh}\over\langle\eta^2\tau_{eddy}\rangle^{2/3}\Omega^{1/3}}
\right].
\end{equation}
As expected, field reversals are exponentially suppressed as we go to
the usual $\alpha-\Omega$ dynamo.
Given $\alpha_{coh}>0$ then as $\alpha_{coh}$ becomes significant we expect
it to enhance the dynamo growth rate and reduce the rate of
spontaneous field reversals.

When $\alpha_{coh}$ is large and negative we can evaluate
equation (\ref{eq:exact})
by integrating the denominator by parts and remembering that the bulk
of the contribution to the integral comes from $w<-1/\gamma$ so that
$w^2\ll-\gamma$.  We obtain
\begin{equation}
\langle A\rangle \approx {\langle\eta^2\tau_{eddy}\rangle\over4\alpha_{coh}},
\end{equation}
and
\begin{equation}
\partial_t\ln B_\theta^2\approx {-3\Omega\langle\eta^2\tau_{eddy}\rangle\over
4\alpha_{coh}}-2D.
\end{equation}
In this limit field reversals occur at a rate given by
\begin{equation}
\tau_{rev}^{-1}=0.78(-\alpha_{coh}\Omega)^{1/2}.
\end{equation}
We note that in this case the coherent component of the helicity
does not completely shut off the incoherent dynamo, even though
by itself it is incapable of driving a dynamo.  Instead we find
that as $|\gamma|$ increases past one the dynamo growth rate
decreases inversely with $|\gamma|$.  Eventually, turbulent diffusion
will suppress the dynamo.  In the limit where $\gamma$
is of order $-1$ we anticipate that the dynamo growth rate will
be less than expected from the incoherent dynamo alone and the
rate of field reversals will be larger.

These analytic results have the advantage of being based on a solvable
model, but do not include the effects
of spatial structure or saturation.  It is therefore instructive
to consider the combined effects of a random electromotive force
and shear in a one dimensional model.  We consider the mean field
equations for a uniform disk with Keplerian rotation and half-thickness
$H$,
\begin{equation}
\partial_t B_r = -\partial_z(\alpha B_\theta)+D_t\partial_z^2B_r,
\end{equation}
\begin{equation}
\partial_t B_\theta = -{3\over 2}\Omega B_r+D_t\partial_z^2B_\theta,
\end{equation}
with $-H\le z\le H$ and $B_r=B_\theta=0$ at $z=\pm H$.
We first consider the incoherent $\alpha$-effect, so we take $\alpha$
to be random in space and time.  When the rms value of $\alpha$ is
large enough we find self-excited solutions that grow without bound.
In reality there must be some quenching mechanism, which we model using
\begin{equation}
\alpha=\alpha_0 {{\cal N}(z,t)\over (1+B_\theta^2)},
\label{eq:br4}
\end{equation}
where ${\cal N}$ is a random function in space and time with zero mean
and an rms value of unity.
Without loss of generality we put $H=\Omega=D_t=1$.

In fig. 1 we plot contours of the $B_\theta$ field in a space-time
diagram for a dynamo number $\alpha_0\Omega H^3/D_t^2$ of $10^4$.
(The critical dynamo number for dynamo action depends on the coherence
time and length scales, $\lambda$ and $\tau$, respectively.  In the
present case we adopt $\lambda=0.05$ and $\tau=0.002$ and find the
critical dynamo number to be around 2000.  At this dynamo number
the ratio of the growth rate given in equation (\ref{eq:gr1}) to $D_t/H^2$
is $\sim 7$.)  The remarkable result is that the $B_\theta$ field
shows a great deal of spatio-temporal coherence with variations
comparable to the diffusion time and diffusion length.  Experiments
with different dynamo numbers suggest that the degree of coherence
is more pronounced for larger dynamo numbers.

In order to isolate the effect of a spatially incoherent $\alpha$-effect
we now investigate a model with a steady $\alpha$-effect of the form
\begin{equation}
\alpha=\alpha_0\sin (n\pi z).
\label{eq:br5}
\end{equation}
{}For large values of $n$, the critical dynamo number is proportional
to $n^2$.  Thus, although the rms value of the $\alpha$-effect is
unchanged, the dynamo becomes harder to excite if $\alpha$ is
chopped into many domains of different sign.  The magnetic
field is steady, and the radial component is of alternating sign.
However, more surprisingly, the toroidal magnetic field has the same
sign for all values of $z$, see fig. 2.  This is very similar to the
simulation of a random incoherent $\alpha$-effect mentioned before.
There is one difference in that the magnetic field shows global reversals
in time when the $\alpha$-effect is incoherent in time.

{}Finally, we note that
in order for the magnetic field to grow the growth rate given in
equation (\ref{eq:gr1}) has to be greater than the dissipation rate.
In general the dissipation rate will depend on the wavenumber of
the magnetic domain as $K^2$, while the growth rate goes as
$(K^2/N)^{1/3}$.  Clearly whether or not there is a self-excited
dynamo will depend in large part on the geometry of the fluid.

\section{The Incoherent Dynamo in Accretion Disks}

A Keplerian accretion disk with a root mean square Alfv\'en speed of $V_A$ will
be subject a local instability first described by Velikhov (1959).
Its pivotal role in transporting angular momentum outward in accretion
disks was
recognized later (\cite{bh91}).  In the context of accretion
disks this instability is normally referred to as the Balbus-Hawley
instability.  Its maximum growth rate is of order $\Omega$, and occurs
at an azimuthal wavelength of $\sim V_A/\Omega$.  In three dimensions
the instability saturates in turbulence with a typical turbulent
velocity comparable to $V_A$ and a typical eddy size of $\sim V_A/\Omega$.
This turbulence is not expected to be isotropic, but the
typical eddies are expected to have axis ratios of order unity,
which in this context means only that no axis should be more
than an order of magnitude larger than another (\cite{vd92}).
Numerical simulations (\cite{bnst95}) indicate that the the azimuthal
scale of the typical eddies is several times the vertical and radial
scales, which is expected in light of the large local shear.
The azimuthal velocity is also larger, although only by a factor of
roughly two.  Neglecting such factors, these scaling laws imply a
turbulent diffusivity of $\sim V_A^2/\Omega$.  The turbulence
largely suppresses the Parker instability and the typical buoyant
velocity of the magnetic field is of order $V_A^2/c_s$ (\cite{vd92},
\cite{v95b}), where $c_s$ is the local sound speed.
The angular momentum flux induced by the turbulence is
approximately $\langle V_\theta V_r\rangle\sim V_A^2$ which implies
a dimensionless viscosity $\alpha_{SS}$ of order $(V_A/c_s)^2$.  Since
$h\Omega\sim c_s$ this implies that magnetic flux is lost from the
disk at a rate which is some fraction of order unity
times $\alpha_{SS}\Omega$.

It is by no means obvious that in real disks this turbulence possesses the
kind of symmetry that would make $\langle \alpha_{\theta\theta}\rangle=0$.
On the other hand, calculations done without vertical structure
or any imposed large scale field (\cite{hgb95b}) give results which
are qualitatively similar to calculations which include vertical
structure (\cite{bnst95}).  By construction
the former calculations must be symmetric under the transformation
$z\rightarrow -z$ even though the latter are not.  We can in principle
estimate $\alpha_{\theta\theta}$ using data from the simulation of
Brandenburg et al. (1995).  From equation (\ref{eq:alpha}) it is
clear that a time integration has to be carried out.  However,
video animations of those data suggest that the life time of
turbulent eddies is shorter than the life time of magnetic structures
which, in turn, is shorter  than the eddy turn over time.  In other
words, the Strouhal number (e.g. Krause \& R\"adler 1980) is small.
As a rough approximation we may therefore replace the time integration
by a multiplication with a relevant time scale.  We adopt the natural
time scale $\Omega^{-1}$, which is sufficient since we are only
interested in relative variations.  We adopt volume averages and
note that because of the periodic boundary conditions in the
toroidal direction,
$\langle V_r V_{z,\theta}\rangle=-\langle V_z V_{r,\theta}\rangle$,
so we can compute
\begin{equation}
\alpha_{\theta\theta}\approx {2\over r}\langle V_r
V_{z,\theta}\rangle\Omega^{-1}.
\end{equation}
In fig. 3 we plot the evolution of $\alpha_{\theta\theta}$ using the
data from run C of Brandenburg et al. (1995), which has now been
carried out for an additional 200 orbits, see also Torkelsson et al. (1996).
This average was computed for the upper half plane of the simulation.
We note that $\alpha_{\theta\theta}$ is positive, in agreement with the
expected effect for bubbles that expand as they rise in a Keplerian disk.
However, the
sign of $\alpha_{\theta\theta}$ suggested by the correlation between the
azimuthal magnetic and electric fields is negative (\cite{bnst95}).
The source of this discrepancy is not yet clear.  In any case the
spatially averaged helicity shows large variations from its long
term average, although the variations in the electromotive force
are much larger.

The size of the fluctuations in the electromotive force, as well as the
persistence of the dynamo in the absence of any $\hat z$ symmetry
breaking, implies that any
preferred helicity resulting from vertical structure is not strong
enough to completely dominate the simulations.  In what follows we will assume
that real disks lack any significant $\langle \alpha_{\theta\theta}\rangle$.
At a minimum our results can be taken as demonstrating that there is an
incoherent dynamo operating in the simulations, and in real accretion disks,
whose effects need to be understood, and cleanly separated from any other
dynamo mechanisms at work.

Let's consider a magnetic domain characterized by the wavenumbers
$(K_r,K_\theta,K_z)$.  Ignoring the anisotropies in the turbulence
we find that the number of turbulent eddies per domain
is roughly $\sim(K_rK_\theta K_z V_A^3)^{-1}\Omega^3$.  Consequently the
growth rate for the dynamo is
\begin{equation}
\tau_{dynamo}^{-1}\sim\left({V_A^5 K_z^3 K_r K_\theta\over
\Omega^2}\right)^{1/3}.
\label{eq:grge}
\end{equation}
However, in a shearing environment we aren't free to specify $K_r$ and
$K_\theta$ separately.  The shear implies a minimum $K_r$ for any
$K_\theta$ since in a time $\sim \tau_{dynamo}$ the shear will increase
$K_r$ by an amount $(3/2)K_\theta\Omega\tau_{dynamo}$.  If we choose a
value of $K_r$ above this minimal value then $\tau_{dynamo}$ will
go as $K_r^{1/3}$ while the dissipation rate scales as $K_r^2$.  Clearly
our chances for a successful dynamo will be maximized by taking
$K_r\sim K_\theta \Omega\tau_{dynamo}$.  This gives us
\begin{equation}
\tau_{dynamo}^{-1}\sim \left({V_A^5 K_z^3K_r^2\over\Omega^3}\right)^{1/2}.
\label{eq:nax1}
\end{equation}
This analysis only makes sense in the limit where the magnetic domains
encompass at least one eddy, or $K_zV_A<\Omega$ and $K_rV_A<\Omega$.
The dissipation rate is roughly
\begin{equation}
\tau^{-1}_{dissipation}\approx (K_z^2+K_r^2){V_A^2\over\Omega}.
\label{eq:diss}
\end{equation}
By comparing equations (\ref{eq:nax1}) and (\ref{eq:diss}) we see
that the incoherent dynamo is incapable of generating non-axisymmetric
large scale magnetic fields.  The dissipation rate of such domains
exceeds the generation rate for all domain sizes greater than a single
eddy.

In a real disk the number of eddies in a magnetic domain does not
increase indefinitely as $K_\theta\rightarrow 0$.  The finite
circumference of the disk implies that for axisymmetric domains
\begin{equation}
N\sim {r\Omega^3\over K_zK_rV_A^3}.
\end{equation}
Consequently, we can rewrite equation(\ref{eq:grge}) as
\begin{equation}
\tau_{dynamo}^{-1}\sim\left({V_A^5 K_z^3 K_r\over r\Omega^2}\right)^{1/3}.
\label{eq:gra1}
\end{equation}
At a fixed wavenumber, and therefore at a fixed dissipation rate,
this rate is maximized for $K_z=K_r3^{1/2}$.  Assuming this ratio
we see that the dynamo growth rate for axisymmetric domains goes
as $K^{4/3}$, which implies that at some sufficiently small $K$
the dynamo will work.  More exactly, the incoherent dynamo caused
by the Balbus-Hawley instability will
drive an increase in the magnetic field strength if
\begin{equation}
K^2<{\Omega\over rV_A}.
\end{equation}
In other words, the incoherent dynamo only works for
\begin{equation}
V_A<{\Omega\over rK^2}.
\label{eq:blim}
\end{equation}

Ultimately $K_z$ is limited by the height of the disk, i.e.
$K_zh>1$.  Moreover, as we approach this limit the buoyant loss of
magnetic flux becomes significant.  The buoyant loss rate from
a single magnetic domain goes as
\begin{equation}
\tau^{-1}_{buoyant}\sim K_z V_b\sim K_z {V_A^2\over c_s},
\end{equation}
so when $K_zh\sim 1$ buoyant losses are as important as turbulent
diffusion.  Of course, the only limit on the radial extent of
a magnetic domain is $K_rr>1$, but lowering $K_r$ past $h^{-1}$
will lower the growth rate without affecting the dissipation
rate.  From equation (\ref{eq:blim}) we see that the magnetic
field associated with scales of order the disk thickness will
be the strongest and will be given by
\begin{equation}
V_A\sim c_s {h\over r}.
\end{equation}
This in turn implies that the dimensionless viscosity associated with
this dynamo mechanism is
\begin{equation}
\alpha_{SS}\sim\left({V_A\over c_s}\right)^2\sim\left({h\over r}\right)^2.
\label{eq:alpha1}
\end{equation}
We expect this scaling law to hold only in the limit $h\ll r$.  As
$V_A\rightarrow c_s$ corrections of order $(V_A/c_s)$ will become
important in our formula for buoyancy.  Since the saturation limit for
the magnetic field involves the small difference between the growth
rate dependence on $V_A$, which has an exponent of $5/3$, and the
buoyant loss rate dependence, which goes as $V_A^2$, we expect the
saturation strength of the magnetic field to be extremely sensitive
to such corrections unless $V_A\ll c_s$.  We also note that
the disk radius enters
into this result only through its role as the circumference of an
annulus.  Computer simulations typically involve a short arc in place
of a full annulus.  In this case the azimuthal length of the simulation
has to be used in place of $r$ in equation (\ref{eq:alpha1}).

The rate of spontaneous magnetic reversals expected in the absence of
any coherent component to $\alpha_{\theta\theta}$ is
comparable to dynamo growth rate.
However, while current simulations seem to show a significant reversal
rate in the presence of vertical structure (\cite{bnst95}), in its
absence the field can evolve for 100 orbital times without
reversing (\cite{tbns96}).  The exact relationship
between the dynamo growth rate and the field reversal rate
dependent on the particular model for the process, and the zero
dimensional model used in this paper may well overestimate the
rate of spontaneous field reversals.  Nevertheless, such reversals
are an intrinsic part of the model, and should occur if the
simulation is run for several growth times.  The sharp rise
in the field reversal rate when vertical structure is included suggests
that a significant coherent $\alpha_{\theta\theta}$
is present in the such simulations.  (In the zero dimensional model
this would argue for a negative vertical gradient in the coherent
$\alpha_{\theta\theta}$.   The situation is less clear in three
spatial dimensions.) In order for this helicity
to allow the buildup of a coherent field in these simulations, as
well as in accretion disks, it has to scale with the local rms
turbulent velocity more steeply than the square of the incoherent
dynamo growth rate, or $(V_T/c_s)^{10/3}$.  If this helicity is
due to the Parker instability and if, as has been argued elsewhere
(\cite{vd92}), the Balbus-Hawley instability reduces the Parker
instability to vertical motions of order $V_T^2/c_s$, then we
can estimate the magnitude of the helicity as
\begin{equation}
V_r k_\theta V_z \tau_{buoyant},
\end{equation}
where $\tau_{buoyant}$ is the correlation time for these buoyant motions.
Since shearing imposes the requirement that
$k_\theta\Omega<k_r\tau_{buoyant}^{-1}$
and since these motions are approximately incompressible, i.e.
$k_r V_r\sim k_z V_z$, this gives a helicity less than
\begin{equation}
V_z^2 k_z/\Omega\sim {V_T^4\over c_s^3}.
\end{equation}
If the coherent helicity has this dependence, then it becomes
important only as the dynamo saturates due to turbulent mixing
and buoyancy.  In this case it will not suppress the incoherent
dynamo in simulations with smaller $(h/r)$, or in real accretion disks,
but it will remain significant in the saturated state.

The fact that the buoyancy does not significantly enhance the loss of
magnetic flux is a critical element in the derivation of equation
(\ref{eq:alpha1}).  Consequently, environments that increase magnetic
buoyancy will saturate at much lower field strengths.  As an example
we can consider magnetic flux tubes in a radiation pressure dominated
environment.  In this case we have (\cite{v95b})
\begin{equation}
V_b\sim {P_{radiation}\over P_{gas}} {V_A^2\over c_s}.
\end{equation}
Combining this result with equation (\ref{eq:gra1}) for
$K_r\sim K_z\sim h^{-1}$ yields
\begin{equation}
\alpha_{SS}\sim\left({V_A\over c_s}\right)^2\sim \left({P_{gas}\over
P_{radiation}}\right)^6\left({h\over r}\right)^2.
\label{eq:alpha2}
\end{equation}

The exponent given in equation (\ref{eq:alpha2}) is perhaps
a bit large in comparison to the value suggested by the
phenomenology of disks, but this is a considerably less
serious problem than if it were too small.  Competing
dynamo mechanisms and/or hydrodynamic angular momentum
transport mechanisms could be driving $\alpha_{SS}$ up.
How does this compare to other sources of viscosity in disks?
The result given in equation (\ref{eq:alpha2}) has an extremely
uncertain coefficient.  Current numerical simulations give
$\alpha_{SS}$ of order $10^{-2}$ or less, which would suggest that
this coefficient is very small.  On the other hand, these
simulations have $V_A\sim c_s$ and are definitely not in the
asymptotic regime where our scaling laws should be valid.
We have already noted that the small difference in the exponent
of $V_A$ in the dynamo growth rate and the dissipation rate,
coupled to the presence
of corrections to both these rates of order $(V_A/c_s)$, makes
it difficult to extrapolate from current results.  If the
saturation value of $V_A/c_s$ approaches its asymptotic dependence
on $h/r$ gradually as $h/r\rightarrow 0$ then the final value
of the coefficient will be much larger than $10^{-2}$.
Bearing in mind the large role that numerical viscosity plays
in the simulations (\cite{v95a}) it seems prudent to regard
the coefficient as an unknown numerical constant.

On the other hand, since $h\ll r$ for many realistic disks
we can compare this dynamo mechanism to others based purely on the
value of the exponent in the scaling relationship.  Of course,
a purely local mechanism {\it coherent} mechanism will not
scale with $(h/r)$ at all, although it might show some dependence
on the local disk temperature.  However, as we noted earlier
this model seems to conflict with phenomenological studies
of dwarf novae and x-ray transients.
Internal waves, excited by tidal instabilities in binary
system disks (\cite{g93}) will produce an effective $\alpha_{SS}$
which scales as $(h/r)^2$ (\cite{vd89}).  This will be
a competing mechanism for angular momentum transport
in gas pressure dominated disks, and potentially
the dominant one in radiation pressure dominated disks.
(Although, such conditions are most likely in AGN disks, where
the potential for the tidal excitation of waves is less certain.)
Given the nonlocal nature of the angular momentum
transport mediated by internal waves, the existence of a
purely local mechanism might be important, even if it
does not clearly dominate.
When the disk is ionized and when internal
waves are present, then the waves are capable of driving a
dynamo with a growth rate $\sim (h/r)^{3/2}\Omega$ (\cite{vjd90})
and perhaps faster, depending on the nature of the turbulent
cascade of wave energy (\cite{vd92}).  The resulting
value of $\alpha_{SS}$ will be $\sim (h/r)^{3/2}(P_{gas}/P)$
(\cite{vd92}, \cite{v95b}).
When these conditions are met this would appear to be
a more important dynamo mechanism, although once again
we note that nonlocal effects on the wave-driven dynamo
make the two processes somewhat incommensurate.  An
equivalent estimate based on purely local physics was given
by Meyer \& Meyer-Hoffmeister (1983).  However,
this estimate is based on using large scale buoyant cells
driven by magnetic buoyancy, a picture which is inconsistent
with turbulence in the disk (\cite{zk75}, \cite{vd92}).
In addition, they assumed approximate isotropy of the
helicity tensor and offered a calculation of $\alpha_{rr}$
instead of $\alpha_{\theta\theta}$. This assumption of isotropy
is inconsistent with the notion that the motions are driven
by magnetic buoyancy, which for $h\ll r$ will have a time
scale much longer than the local shearing time scale.
{}Finally, we note that our result for the incoherent dynamo
in an accretion disk is sensitive to our assumption that
the process must be self-exciting.  Given an external source
of fluid turbulence, e.g. convection or the turbulent cascade
of energy set off by finite-amplitude internal waves, the
incoherent dynamo may have a significantly larger growth rate
and give rise to a larger $\alpha_{SS}$.

\section{Conclusions}

In this paper we have shown that mean-field dynamo
theory allows for the existence of a new kind of
$\alpha-\Omega$ dynamo, which we have named the
incoherent $\alpha-\Omega$ dynamo, in which there is no coherent
helicity whatsoever.  In this class of dynamos the
magnetic field is driven by a combination of a random
walk for $B_r$ and its shearing, which creates $B_\theta$.
The resultant large scale field derives its organization from
coherent shearing effects, rather than any loss of mirror
symmetry in the turbulence.  Although this kind of dynamo
necessarily includes spontaneous field reversals, such reversals
may occur at a rate
which is some fraction of the dynamo growth rate.  The existence
of a mean-field dynamo in a flow with a mean helicity of
zero is interesting for its own sake, since it provides an
example of how large scale order in the magnetic field can
arise from the interaction between a large scale shear and
statistically symmetric local motions.  In this sense it
represents an alternative to models which seek to explain
dynamo activity through asymmetric turbulence and a coherent
helicity.  It differs from previous attempts to do without a
coherent helicity (e.g. Montgomery et al. (1984) and Gilbert
et al. (1988)) in that it does so without appealing to
the other terms in equation (\ref{eq:dyn1}).
This model is particular interesting in light of previous
claims (\cite{m79}) that the coherent $\alpha$ effect does not
converge (although see Kraichnan (1979) for a counterargument).

This dynamo is particularly interesting in light of simulations
of magnetic field instabilities in accretion disks.  We have
suggested that this dynamo can operate successfully in accretion
disks, but only
to produce axisymmetric large scale fields.  Comparing the
growth rate for this dynamo with the buoyant loss rate for
magnetic flux, we see that {\it if} this is the only dynamo
associated with magnetic shearing instabilities, then the
large scale magnetic field will saturate when $V_A\sim (h/r)c_s$
and $\alpha_{SS}\sim (H/r)^2$.  This result may seem somewhat
odd, since the dynamic equations do not depend on $r$ at all.
However, the factor of $r$ comes in through geometrical considerations,
i.e.  from considering the number of independent
eddies in an axisymmetric magnetic field domain. In that
sense it refers
to the circumference of such an annulus rather than its radius.
Consequently, when comparing numerical simulations to the
predictions of this model one should substitute the azimuthal
extent of the simulations for $r$.  For current simulations
this gives $(h/r)\sim (2\pi)^{-1}$.   In fact, since the
turbulent eddies are longer in the azimuthal direction, the number
of independent eddies that can be stacked end to end in current
simulations is one or two, so that effectively $(h/r)\sim 1$.
A critical test of this model would be to run simulations where
this ratio was small and look for a drop in $V_A/c_s$ in the
saturated state.  It's interesting to note that the only numerical
simulation with no imposed field and with disk vertical structure
does seem to have a coherent component to the helicity, which
may have the wrong sign to drive a conventional dynamo.  This may account
for the large rate of spontaneous field
reversals when vertical structure is put in the simulations.
The existence of such a component is consistent with the existence
of an incoherent $\alpha-\Omega$ dynamo, but only if its amplitude
scales steeply with the strength of turbulence in the disk.

{}Finally, we note that this model successfully reconciles
phenomenological models of stellar accretion disks and
the existence of a dynamo effect in a magnetized disk.
The only drawback is that this model gives a relationship
between $\alpha_{SS}$ and $(h/r)$ which is probably too steep,
implying the existence of other, more efficient dynamo mechanisms
in accretion disks in binary systems.

\acknowledgements
This work has been supported in part by NASA grant NAG5-2773.
In addition, ETV would like to thank the Harvard-Smithsonian
Center for Astrophysics for their hospitality while this paper
was being written.

\clearpage
\begin{figure}
\plotone{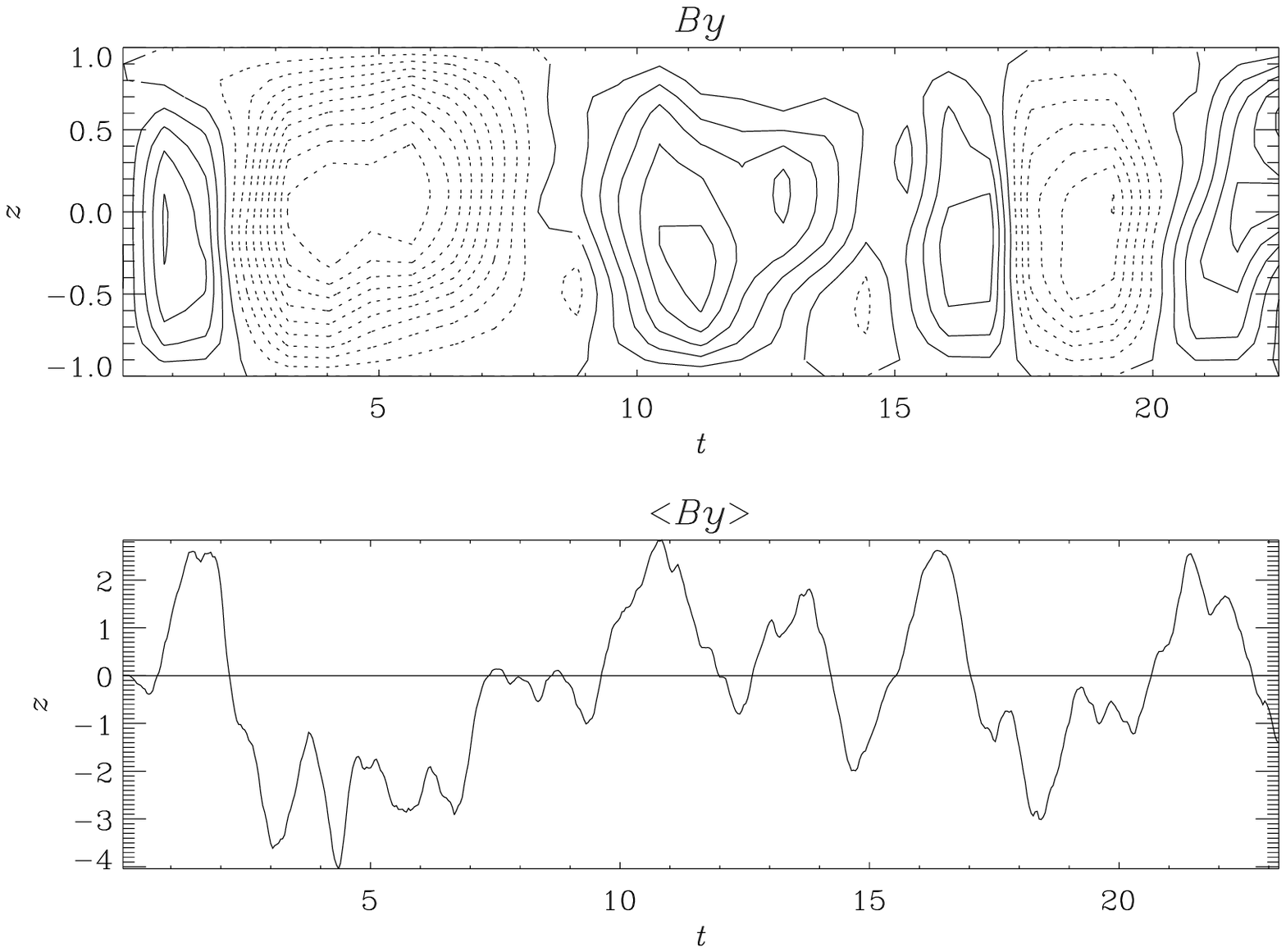}
\caption{Contours of the $B_\theta$ field in a space-time diagram for
the one dimensional spatially and temporally incoherent dynamo model.}
\end{figure}
\begin{figure}
\plotone{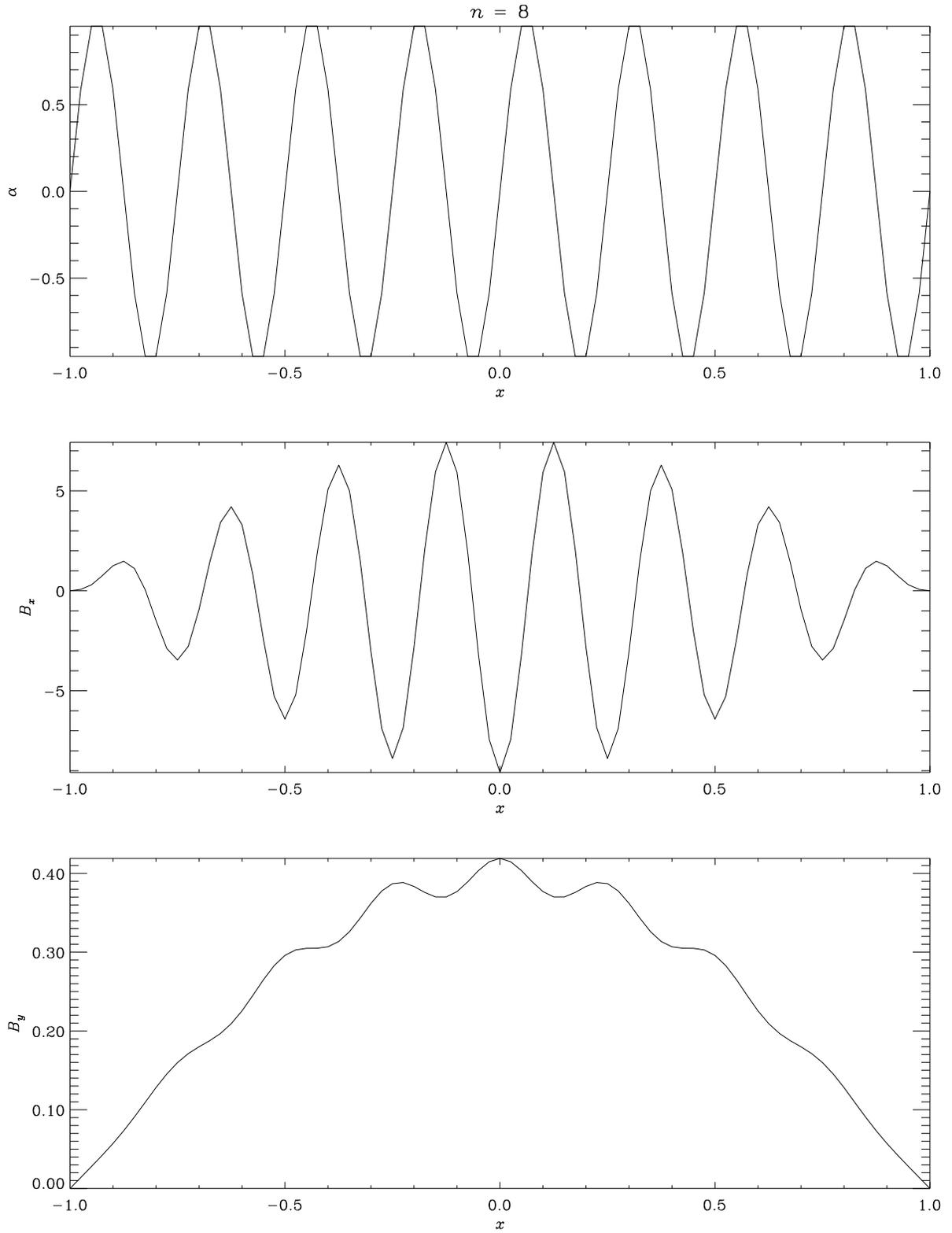}
\caption{A snapshot of the magnetic field and helicity for
the one dimensional spatially incoherent dynamo model.}
\end{figure}
\begin{figure}
\plotone{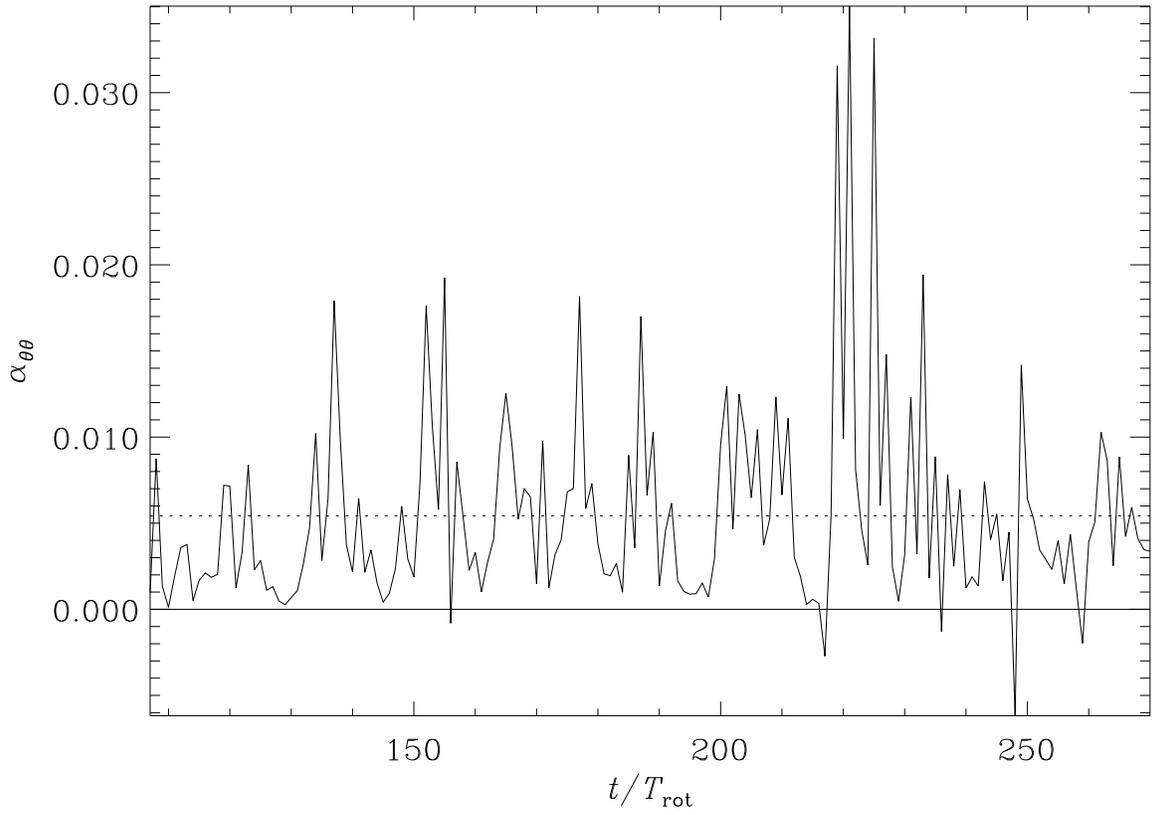}
\caption{The evolution of $\alpha_{\theta\theta}$ (normalized by the
product of the rms velocity and $(\Omega\tau_{eddy})^{-1}$) in the upper half
plane of run C from Brandenburg et al. (1995).}
\end{figure}
\end{document}